# Synthesis by molten salt method of the AFeO$_3$ system (A = La, Gd) and its structural, vibrational and internal hyperfine magnetic field characterization

M. Romero[1], V. Marquina[1], R.W. Gómez[1], J.L. Pérez-Mazariego[1] and R. Escamilla[2]

[1]Facultad de Ciencias, Universidad Nacional Autónoma de México, Apartado Postal 70-399, México D. F., 04510.
[2]Instituto de Investigaciones en Materiales, Universidad Nacional Autónoma de México, Apartado Postal 70-360, México D. F., 04510

**ABSTRACT.** Polycrystalline samples of LaFeO$_3$ and GdFeO$_3$ were synthesized by the molten salt method. Some properties and the quality of the resulting compounds were investigated. The crystal structure and purity of the samples was determined through X-ray diffraction and Rietveld analysis. The vibrational properties were characterized by Raman and IR spectroscopy. Mössbauer spectroscopy was used to determine the ionic state of the Fe ions and the internal hyperfine magnetic fields Considerable reduction of the heat treatment (temperature and time) for the reaction to take place was achieved without detriment of the quality of the compounds.

## 1. INTRODUCTION

It is known that natural and synthetic perovskites can attain a wide variety of electronic properties that comprise insulators, semiconductors, conductors and high $T_c$ superconductors, depending on their composition and structure. They also show several magnetic properties, such as ferromagnetism, anti-ferromagnetism, magnetoresistance, ferroelectricity, etc. and much research has been done in these ceramic materials.
The general chemical formula of perovskites is $ABX_3$, where A and B are metallic cations and X is a non-metallic anion [1]. Ideally, they have a cubic structure in which cations A occupy the eight corners of the cubes and the smaller cations B their centers; the X anions are face centered. However, the different sizes of the A ions can distort the cubic structure in several ways diminishing its symmetry [2].

Among the distorted perovskites, the (La,Gd)FeO$_3$ system acquires an orthorhombic structure, with a Pbnm or Pnma space group [3], due to the tilting of the [FeO$_6$] octahedral (Figure 1). The potential technological applications of these orthoferrites has raised much interest in relation to the properties mentioned in the preceding paragraph, such as magnetic field sensors (magneto-resistance), actuators (piezoelectricity) and oxygen sensors based in their electronic-ionic mixed conductivity with nonlinear response to oxygen pressure [4-10].

They also show feeble antiferromagnetism and the LaFeO$_3$ has the highest Néel temperature (738 K) of the orthoferrite family [11].

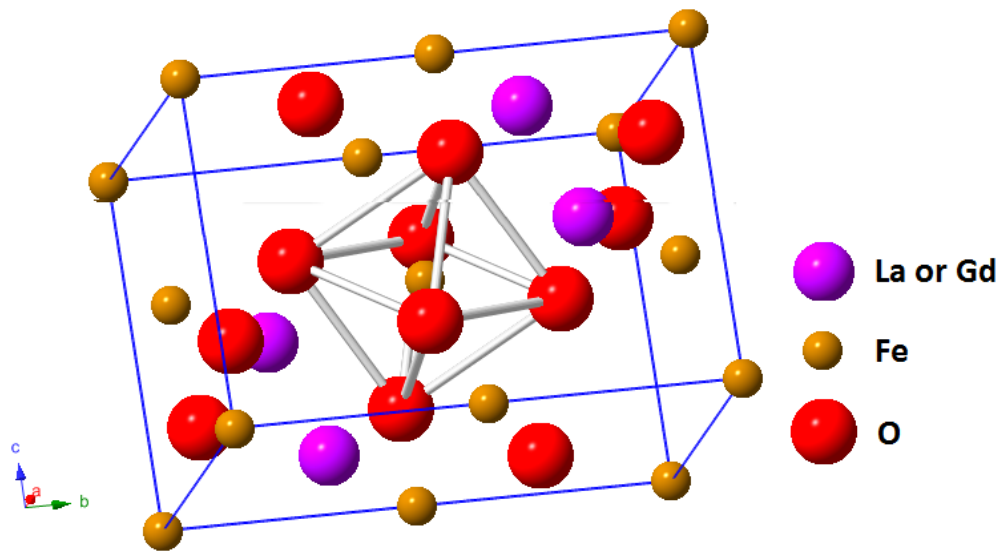

Figure 1. Structure of the (La, Gd)FeO$_3$ system

In this work we carry out an extensive characterization of both compounds synthesized using the molten salts procedure to explore its advantages against other synthesis methods, such as co-precipitation [12, 13], combustion [14], solid-state reaction [15, 16] and sol-gel [17]. Their characterization was done by X-ray diffraction (XRD), infrared (IR), Raman (RS) and Mössbauer (MS) spectroscopies.

2. **Material and methods**

To prepare polycrystalline samples of both LaFeO$_3$ and GdFeO$_3$, mixtures of KCl (99.0%) and NaCl (99.5%) Sigma-Aldrich, were first grounded separately and then together until a fine powder was obtained. The same procedure was employed with a stoichiometric mixture of Fe$_2$O$_3$ (99.0 %), and La$_2$O$_3$ or Gd$_2$O$_3$, respectively; finally, a grounded 1:10 molar reactants to salt proportion mixture was 6 hours heat treated at 900 °C. The melting temperature of the eutectic salt mixture is 650 °C [18, 19] so its purpose is to serve as a liquid medium to improve the mobility of the reactants to favor their reaction times.

Room temperature X-ray 2θ powder diffraction patterns were obtained with a Siemens (D5000) diffractometer in 0.02° steps from 20 °-120 ° using the Co-K$_\alpha$ radiation Ni filtered; cell parameters were Rietveld refined with the least squares MAUD program [20].

A attenuated total reflectance (Thermo Scientific, model Smart Orbit) accessory attached to a FT-IR Infrared Thermo Scientific, model Nicolet 6700, was used to obtain the IR spectrum of the sample in powder form, after determining the analytical background.
Raman-scattering measurements were performed with a Horiba Jobin Yvon Xplora Olympus BX41 spectrometer equipped with an optical microscope and a photomultiplier tube in a back scattering geometry. A 40× objective was adopted to focus the laser beam of a class 3B laser (532 nm) continuous wave with a power of 25 mW. Different areas of the samples were measured in order to ensure average sample inhomogeneities. Convolution of the observed bands was done with a Gauss-Lorentz oscillator model. Fitted curves showed less than 1% error with the corresponding spectra.

Thin absorbers, prepared from the powder samples, were used to record room temperature spectra in a constant acceleration transmission Mössbauer spectrometer, using a $^{57}$Co Mössbauer source in Rh matrix. All spectra were fitted with the Recoil 1.05 program [21]. The reported isomer shifts are respect to α-iron (Table 4)

### 3. Results

#### 3.1 X-Ray Diffractograms

Two samples, one of LaFeO$_3$ and one of GdFeO$_3$, were grounded and the obtained powder was used to generate an analysis of the crystal structure. The recorded diffraction patterns are shown in Figure 2. The comparison of the X-ray powder diffraction pattern of the LaFeO$_3$ compound (Figure 3a) with the corresponding ICSD 28255 card does not show the presence of impurities, whereas in the GdFeO$_3$ (Figure 3b) a small amount of Gd$_2$O$_3$ is revealed after comparison with ICSD 16644 and ICSD 27996 cards. In both cases the Rietveld refined parameters (Table I), are comparable with previously reported ones in the literature [22].

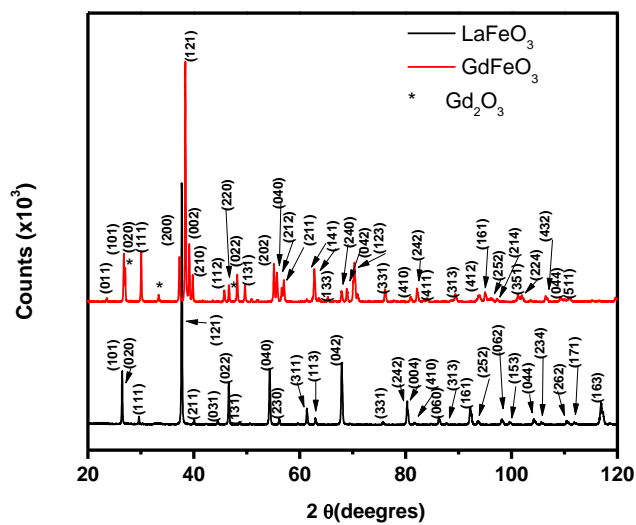

Figure 2. X-ray spectra of the LaFeO$_3$ and GdFeO$_3$ samples. Through the Miller indices the crystal structure of each compound and the impurity (Gd$_2$O$_3$) are identified.

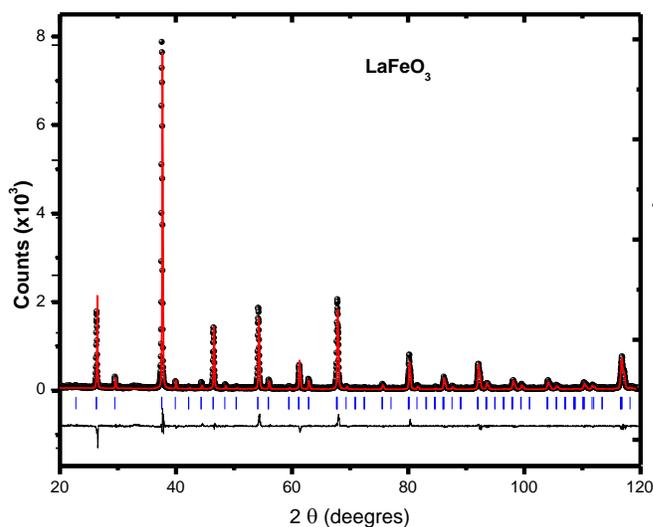 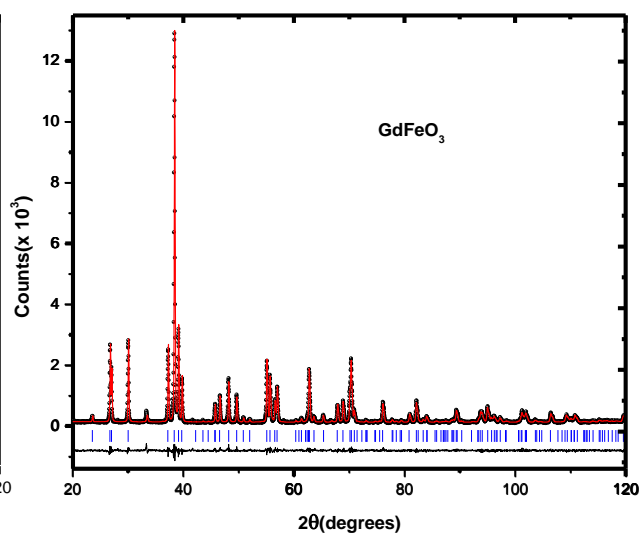

| Figure 3a) | Figure 3b) |

Figure 3. X-ray difraction paterns of the compounds (black dots) and their corresponding Rietveld refinements (red line). a) LaFeO$_3$ and b) GdFeO$_3$. The blue vertical lines are the positions of the reflections reported in the database. The bottom black line is the difference between the experimental and calculated data.

Table 1. Lattice parameters, phase percentage and cell parameters obtained from the Rietvel refinement of LaFeO$_3$ and GdFeO$_3$ compounds.

| Lattice parameters (Å) | LaFeO$_3$ | GdFeO$_3$ |
|---|---|---|
| a | 5.5676(2) | 5.6094(7) |
| b | 7.8608(3) | 7.6720(1) |
| c | 5.5596(2) | 5.3523(7) |
| % de phase in sample | 100 | 98.71 |
| Gd$_2$O$_3$ |  | 1.29 |
| $\chi^2$ | 1.29 | 1.22 |
| R$_w$ | 14.59 | 7.99 |
| R$_b$ | 12.56 | 6.07 |
| R$_{exp}$ | 11.27 | 6.53 |

At room temperature, all samples undertake a pseudocubic structure of orthorhombic symmetry described by the $D_{2h}^{16}$ space group Pnma or Pbnm setting. We report the (La, Gd)FeO$_3$ compounds with orthorhombic phase with Pnma space group (62). Whereas the Fe-O$_1$ bond lengths of both compounds are nearly the same, there is a noticeably shortening of the Fe-O$_2$ bond lengths in the La (1.736 Å) compound respect to the Gd one (2.013 Å) associated to the different ionic radii[†] (Table 2). In consequence the Fe-O$_2$-Fe bond angles of AFeO$_3$ increase from 168.53° for La, to 147.83° for Gd, as determined from the Rietveld

---

[†]The ionic radii of La and Gd in eight-fold coordination are 1.16 Å and 1.053 Å respectively

analysis of X-ray diffraction data, with a concomitant orthorhombic distortion due to the tilting of the Fe-O distorted octahedra. The tilt angle α between the octahedra along the *c*-axis is bigger for GdFeO$_3$ than for LaFeO$_3$ and the distortion of the structure is smaller.

Table 2. Bond lengths in the crystal structure of LaFeO$_3$ and GdFeO$_3$.

| Bond length (Å) | GdFeO$_3$ | Bond length (Å) | LaFeO$_3$ |
|---|---|---|---|
| Gd-O1 | 2.296(2) | La-O1 | 2.267(7) |
| Fe-O1 | 1.998(3) | Fe-O1 | 2.043(1) |
| Fe-O2 | 2.013(2) | Fe-O2 | 1.736(7) |
| Fe-Gd | 2.345(3) | Fe-La | 2.266(1) |

3.2 Raman Spectroscopy

Based in the symmetry operations of the Pnma space group carried out by Smirnova et al [23] the following assignment of the bands shown in Figure 4 are made. Modes caused by La and Gd vibrations are present below 200 cm$^{-1}$, labeled (A). Modes between 200 and 300 cm$^{-1}$ are oxygen octahedral tilt modes (*T*) in La, and 200 cm$^{-1}$ and 400 cm$^{-1}$ in Gd. Modes between 400 cm$^{-1}$ and 450 cm$^{-1}$ are oxygen octahedral bending vibrations (*B*) and modes above 500 cm$^{-1}$ are oxygen stretching vibrations (*S*) [24–26, 27, 28]. In Figure 5 the low energy bands for both compounds are shown up to 600 cm$^{-1}$, labeled according to the $A_g$ and $B_{2g}$ symmetry of the Pnma space group. All the bands below 300 cm$^{-1}$ in the La compound are shifted towards higher energies respect to the Gd ones. The differences between both spectra are associated with the dissimilar distortions caused by La and Gd.

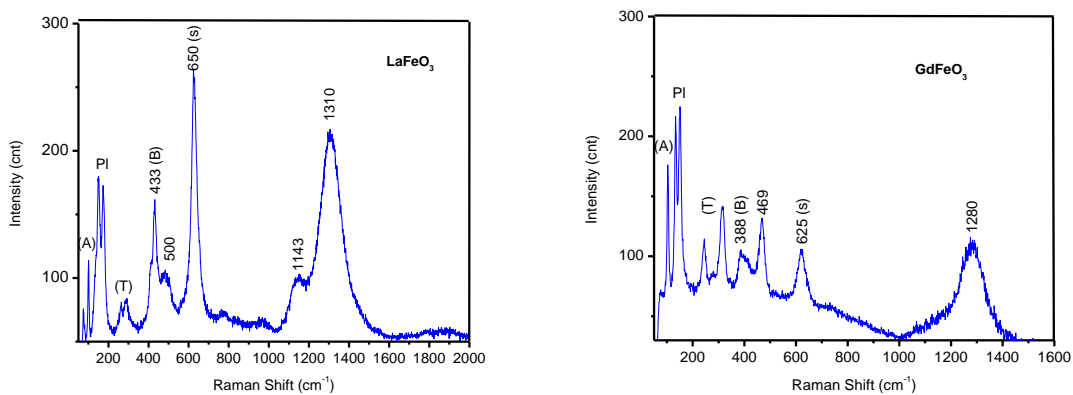

Figure 4. Raman spectra of the LaFeO$_3$ and GdFeO$_3$ compounds.

In particular, the band around 1310 cm$^{-1}$ in the La compound and the one around 1280 cm$^{-1}$ in the Gd can be endorsed to second-order excitation of the 650 cm$^{-1}$ band in LaFeO$_3$ and the 625 cm$^{-1}$ band in GaFeO$_3$. The 1143 cm$^{-1}$ in the La compound could be due to a second-order excitation of the mixture of the $B_{2g}$ excitations 2(151 + 431) cm$^{-1}$ = 1164 cm$^{-1}$, In the Gd compound, however, the width of the corresponding 2(133+412) cm$^{-1}$ =1090 cm$^{-1}$ and 1280 cm$^{-1}$ bands impedes their resolution.

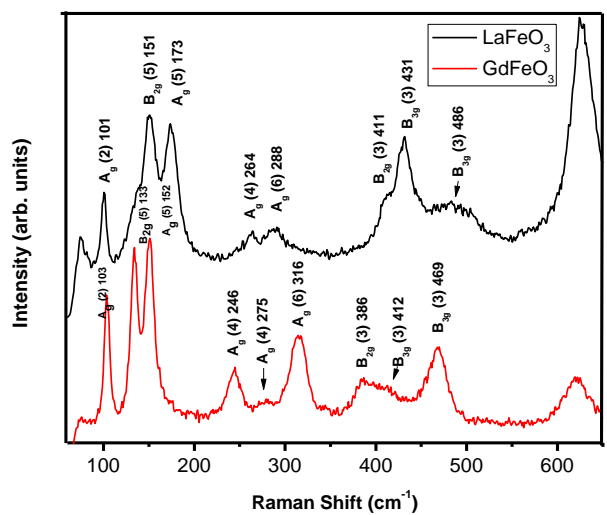

Figure 5. Comparison of the Raman spectra at frequencies below 600cm$^{-1}$ for GdFeO$_3$ LaFeO$_3$ and compounds.

3.3 Infrared Spectroscopy

The Pbnm or Pnma phases should have 25 dipole-active optical phonon modes, 9B$_{1u}$ + 7B$_{2u}$ + 9B$_{3u}$ that rage from around 115 cm$^{-1}$ to 645 cm$^{-1}$ [23]; however, only 9 fall in the (400-700) cm$^{-1}$ range. In Figure 7 nine bands are present in the IR absorbance spectrum of the La compound, that are very close to the calculated ones (Table 3), but line broadening allows to distinguish only five in the Gd compound.

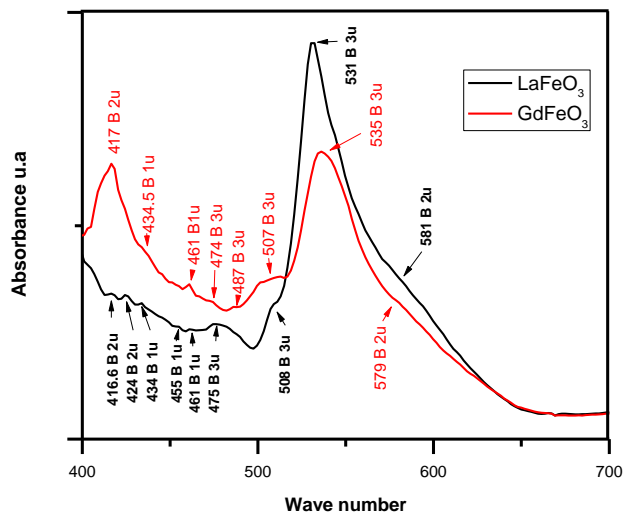

Figure 6. IR-spectra obtained experimentally for LaFeO$_3$ and GdFeO$_3$ compounds at room temperature

Table 3. Calculated and experimental transverse (TO) and longitudinal (LO) frequencies (cm$^{-1}$) of IR-spectra phonon modes ω for LaFeO$_3$ and GdFeO$_3$ compounds.

| Pnma Phase Ref. [29] | | LaFeO$_3$ This Work | GdFeO$_3$ This Work |
|---|---|---|---|
| ω$_{TO}$ | ω$_{LO}$ | | |
| 420 B 2u | 426 B 2u | 416(424) | 417 |
| 434 B 1u | 450 B 1u | 434 | 434.5 |
| 455 B 1u | 457 B 1u | 455(461) | 461(474) |
| 473 B 3u | 479 B 3u | 475(508) | 487(507) |
| 528 B 3u | 531 B 3u | 531 | 535 |
| 573 B 2u | 598 B 2u | 581 | 579 |
| 634 B 2u | 640 B 2u | | |

### 3.4 Mössbauer Spectroscopy

The Mössbauer spectra of both compounds are shown in Figure 7 and their parameters appear in table 4. In both cases, they consist of a single sextet with essentially the same isomer shift (IS), characteristic of Fe$^{3+}$ in high spin. The quadrupole splitting (ΔQ) for the Gd compound is slightly smaller than that for the La compound, with magnitudes close to zero, reflecting the fact that the oxygen octahedra around the Fe ions are almost regular. The smaller value of ΔQ for the former is to be related with the bigger Fe-Gd bond length respect to Fe-La. In addition, as expected from the X-ray results, the hyperfine field magnitude from Gd is slightly smaller than the one for La, due to the less distorted structure of the former.

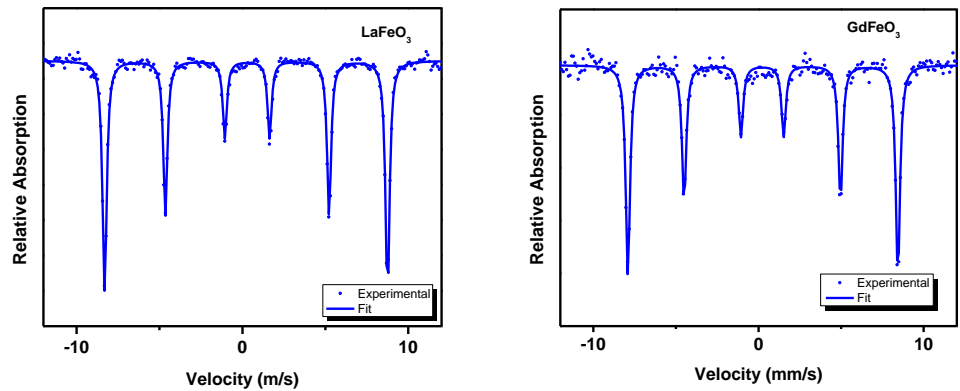

Figure 7. Mossbauer spectra of the samples.

Table 4. Parameters Mossbauer for LaFeO$_3$ and GdFeO$_3$ compounds

| SAMPLE | IS (mm/s) | ΔQ (mm/s) | H (T) |
|---|---|---|---|
| LaFeO$_3$ | 0.38±0.003 | 0.03±0.003 | 52.90±0.02 |
| GdFeO$_3$ | 0.37±0.005 | 0.02±0.005 | 50.70±0.03 |

4. Discussion

The several characterizations used in this work corroborate the quality of the samples obtained by the molten salt reaction route, exhibiting its advantages over the more common solid-state reaction route, that can take as long as 12 days of heat treatment at as high as 1473 K (1200 °C). The heat treatment time and temperature for the reaction to take place are substantially reduced in our method, with no detriment on the purity of the products, as judged from the X-ray diffraction, IR and Raman results.

The increasing orthorhombic distortion of AFeO$_3$ (A=La and Gd), associated with the different ionic radii of La (0.116 nm) and Gd (0.1053 nm) in eightfold coordination, is manifested by a decrease in the Fe-O$_2$-Fe bond angles from 168.53° for A=La, to 147.83° for A=Gd, as determined from Rietveld analysis and X-ray diffraction data.

The smaller hyperfine field of Gd respect to La must be related with the tilt angle and Néel temperature dependence with the ionic radius dimensions of the A observed in [30-33], the former and by [6, 8 and 17] the later.

5. Conclusions

Lower temperatures and reaction times, compared to solid-state reactions, were needed to synthesize pure AFeO$_3$ (A=La and Gd) orthoferrites via molten salt reaction. Their characterization using X ray diffraction, IR, Raman and Mössbauer spectroscopies confirm a pure phase in the case of A = La and an almost pure phase for A = Gd. The small differences in the quadrupole splitting and hyperfine fields observed in the Mössbauer spectra are consistent with the bond lengths and tilt angles differences measured with XRD.

Acknowledgments

This work was partially support by the project IN 115612, DGAPA, UNAM and the Programa de Becas Posdoctorales de la UNAM, DGAPA, UNAM. The authors want to thanks F. Barffuson and M. Canseco for their technical support.